\documentclass[11pt,a4paper]{JHEP}
\usepackage{epsfig,bbm}

\def\bX{\overline X}
\def\A{{\cal A}}
\def\F{{\cal F}}
\def\Z{{\mathbbm Z}}
\def\d{\partial}
\def\dw{\partial_w}
\def\dc{\partial_\zeta}
\def\dbw{\partial_{\bar w}}
\def\dbc{\partial_{\bar \zeta}}
\def\Tr{{\rm Tr}}
\newcommand\0{\nonumber}
\newcommand\ee{\end{eqnarray}}	 	%eqnarray
\newcommand\be{\begin{eqnarray}}
\newcommand\ba{\begin{array}}			%array
\newcommand\ea{\end{array}}
\newcommand\bma{\left(\ba{cccccccc}}
\newcommand\ema{\ea\right)}			%matrix
\newcommand\e{{\rm e}}
\def\o{{\omega}}
\newcommand\cH{{\cal H}}
\newcommand\cP{{\cal P}}
\newcommand\Lm{{\Lambda}}

\preprint{SISSA/45/98/EP\\hep-th/9805071}

\title{Matrix String Theory, 2D SYM Instantons and affine Toda systems}

\author{G.\ Bonelli, L.\ Bonora, F.\ Nesti\\
International School for Advanced Studies (SISSA/ISAS)\\
Via Beirut 2--4, 34014 Trieste, Italy, and INFN, Sezione di Trieste\\
E-mail: \email{bonelli@sissa.it}, \email{bonora@frodo.sissa.it}, 
\email{nesti@frodo.sissa.it}}

\abstract{Extending a recent result of S.B.\ Giddings, F.\ Hacquebord and 
H.\ Verlinde, we show that in the $U(N)$ SYM Matrix theory  there exist classical 
BPS instantons which interpolate between different closed string configurations
via joining/splitting interactions similar to those of string field theory.
We construct them starting from branched coverings of Riemann surfaces.
For the class of them which we analyze in detail the construction can be 
made explicit in terms $U(N)$ affine Toda field theories.\\\\
{\sc Pacs:} 11.15.-9, 11.25.Sq, 11.27.+d}


\begin{document}

\section{Introduction}

The ${\cal N}=(8,8)$ $U(N)$ SYM field theory in 1+1 dimensions is believed to 
represent, in the
strong coupling limit, a theory of closed superstrings \cite{motl,BS}. The
relevant action can be obtained from M(atrix) theory, \cite{BFSS}, via 
compactification on a circle~\cite{WT}. 
More precisely, it has been suggested that this theory describes a second 
quantized superstring theory~\cite{DVV} (see also~\cite{DMVV,HV,BC} and the 
review article~\cite{DVVr}). To confirm this attractive  
conjecture one should be able to extract some evidence from the very structure
of the SYM theory in a region of strong coupling, corresponding to weak string 
coupling, where the expected behavior of the theory should correspond to
perturbative string field theory.

Considerable progress in this direction has been made recently in~\cite{GHV}, 
where the authors show the existence of a 2D instanton which 
interpolates between two string configurations and exhibits the typical
joining/splitting interaction of strings. The idea is that this and other
similar instantons generate a quantum tunneling between given initial and 
final string configurations. In this paper we would like to 
proceed along the same line of investigation and show that many other 2D 
instantons exist which are relevant to string interactions, and outline a
possible classification of them. 

The outcome of our analysis can be summarized as follows.
In the $U(N)$ SYM Matrix string theory (henceforth, simply MST) there exist 
classical BPS instantons which
interpolate between different closed string configurations via  
joining/splitting interactions similar to those of string field theory. Such 
instantons can be constructed starting from (branched coverings
of) oriented Riemann surfaces with punctures or boundaries.
In general they correspond to Hitchin systems on the cylinder.
For the class of them (corresponding to $\Z_n$ coverings), which 
we have studied in detail, it is possible to give a 
rather explicit construction in terms of classical solutions of the affine 
$U(N)$ Toda field theory.

These results provide further evidence that
MST becomes, in the strong coupling limit, a theory of closed (super)strings. 
They tell us that $U(N)$ SYM theory
can describe string interactions and show how
string world--sheet Riemann surfaces make their appearance in MST. 

The paper is organized as follows. In section~\ref{sec:2} we identify the equations 
satisfied by classical SYM configurations that preserve half supersymmetry. 
They define Hitchin 
systems~\cite{hitchin}. In order 
to find explicit solutions we follow \mbox{\cite{wynter,GHV}} and dress 
the singularities corresponding to the string interactions
by means of suitable singular gauge transformations. This allows us,
in particular, as mentioned above, to expose
the structure of an affine $U(N)$ Toda field theory system underlying a class of
instanton solutions (section~\ref{sec:3}). Section~\ref{sec:4} is
devoted a discussion of these results. An Appendix deals with the
particular case of the $\Z_3$ covering.

\section{The 2d SYM model and classical supersymmetric configurations}
\label{sec:2}

\subsection{Minkowski version}\label{sec:21}

The $U(N)$ SYM model in a 1+1 Minkowski space is specified by the action
\begin{equation} 
S = -\frac{1}{2\pi} \int d\sigma d\tau\,\Tr \left(
D_\mu X^i D^\mu X^i + \frac{1}{2g^2} F_{\mu \nu}F^{\mu\nu}-
\frac{g^2}{2}[X^i,X^j]^2 -i \bar\theta \rho^\mu  D_\mu \theta 
- g \theta^T \Gamma_i [X^i,\theta] \right),\label{mSYM}
\end{equation}
where $g$ is the gauge coupling, $\sigma$ and $\tau$ are the world--sheet
coordinate on the cylinder. 
In this equation $\mu,\nu=0,1$, and the 2D flat Minkowski metric 
$\eta_{\mu\nu}$ is taken to have signature $(-,+)$. $X^i$ with $i=1,\ldots,8$ 
are hermitean $N\times N$ matrices and
$D_\mu X^i = \partial_\mu X^i + i[A_\mu, X^i]$. $F_{\mu\nu}$ is the curvature
of $A_\mu$. Summation over the $i,j$ indices is understood. 
$\theta$ represents 8 $N\times N$ matrices whose entries are 2D 
spinors. It can be written as $\theta^T= (\theta^-_s,\theta^+_c)$, 
where $\pm$ denotes the 2D chirality and $\theta^-_s,\theta^+_c$ 
are spinors in the ${\bf 8_s}$ and ${\bf 8_c}$ representations of $SO(8)$, 
while $^T$ represents the 2D transposition.  $\rho_\mu$ are the 2D gamma 
matrices: $\{\rho_\mu, \rho_\nu \}= - 2 \eta_{\mu\nu}$, and $\bar \theta = 
\theta^T \rho^0$. The matrices $\Gamma_i$ are the $16\times 16$ $SO(8)$ 
gamma matrices. For definiteness we will write the matrices $\rho_\mu$ and 
$\Gamma_i$ in the form
\begin{equation}
\rho^0 = \left(\matrix{0& -i\cr i&0}\right),
\quad\quad \rho^1 = \left(\matrix{0& i\cr i&0}\right),\quad\quad
\Gamma_i = \left(\matrix {0 & \gamma_i\cr
                 \tilde \gamma_i & 0\cr} \right),\0
\end{equation}
and $\gamma_i,\tilde\gamma_i$ are the same as in Appendix 5B of~\cite{GSW}.

The action (\ref{mSYM}) is invariant under the supersymmetric transformations
\begin{eqnarray}
&&\delta X^i = \frac{i}{g} \epsilon \Gamma^i \theta\0\\
&&\delta \theta = \frac{1}{2g^2} \rho^{\mu\nu} F_{\mu\nu} \epsilon -
\frac {i}{2} [X^i,X^j]\Gamma_{ij} \epsilon -\frac{1}{g} \rho^\mu D_\mu X^i
\rho^0 \Gamma_i\epsilon\0\\
&& \delta A_\mu = -i \bar \epsilon \rho_\mu \theta\,,\label{msusy}
\end{eqnarray}              
where $\epsilon^T = (\epsilon^-,\epsilon^+)$ are the 8+8 transformation 
parameters.

\subsection{Euclidean version}\label{sec:22}

We make a Wick rotation and introduce the complex coordinates 
\begin{eqnarray}
w= \frac {1}{2} (\tau +i \sigma),\quad \bar w = \frac {1}{2} (\tau - i \sigma),
\quad\quad A_w= A_0-iA_1 ,\quad A_{\bar w}= A_0+iA_1\,.\0
\end{eqnarray}
The action becomes
\begin{eqnarray} 
S&=&\frac{1}{\pi} \int d^2w \,\Tr \left(
D_w X^i D_{\bar w} X^i - \frac{1}{4g^2} F_{w\bar w}^2 -
\frac{g^2}{2}[X^i,X^j]^2\right. \0\\
&&~~~~~~~~~~~~~~~~~~~\left. +i (\theta^-_s D_{\bar w} \theta^-_s + \theta^+_c
D_w \theta^+_c) + ig \theta^T \Gamma_i [X^i,\theta] \right),\label{eSYM}
\end{eqnarray}
The supersymmetry transformations take the form
\begin{eqnarray}
&&\delta X^i = \frac{i}{g} (\epsilon^-_s \gamma^i \theta^+_c +
\epsilon^+_c \tilde\gamma^i \theta^-_s)\0\\
&&\delta \theta^-_s = (-\frac{i}{2g^2} F_{w\bar w}  +
\frac {1}{2} [X^i,X^j]\gamma_{ij}) \epsilon^-_s -\frac{1}{g}D_w X^i
\gamma_i\epsilon^+_c\0\\
&&\delta \theta^+_c = (\frac{i}{2g^2} F_{w\bar w} +
\frac {1}{2} [X^i,X^j]\tilde\gamma_{ij}) \epsilon^+_c 
-\frac{1}{g}  D_{\bar w} X^i \tilde\gamma_i\epsilon^-_s\0\\
&& \delta A_w = -2\epsilon^-_s \theta^-_s, \quad\quad \delta A_{\bar w}=
-2\epsilon^+_c \theta^+_c\,,\label{esusy}
\end{eqnarray}              
where
\begin{eqnarray}
\gamma_{ij} = \frac {1}{2} (\gamma_i\tilde\gamma_j - \gamma_j\tilde \gamma_i),
\quad\quad \tilde\gamma_{ij} = \frac {1}{2} (\tilde\gamma_i\gamma_j - 
\tilde\gamma_j\gamma_i)\,.\0
\end{eqnarray}
In the following we consider various coordinate transformations; in particular,
a string interpretation is most natural after the coordinate transformation
$w \to z= \e^w$, i.e.\ after passing from the cylinder to the annulus or
the complex plane. If we make such transformation on the action (\ref{eSYM}),
conformal factors are bound to appear in the second, third and last terms
of (\ref{eSYM}), since they are not conformal invariant.

\subsection{The string interpretation}\label{sec:23}

For later use let us briefly recall how the string interpretation arises
in the above theory. The naive strong coupling limit ($g\to \infty$) in 
the action tells us that all the $X^i$ and $\theta$ commute, therefore 
they can be simultaneously diagonalized, and the theory becomes a free theory
of the diagonal degrees of freedom, with a residual gauge freedom reduced to 
the Weyl group. The latter in turn can be interpreted as free strings of 
various lengths. For example, let $\hat X^i = Diag(x_1^i,\ldots,x_N^i)$ and let us 
consider the effect on such configuration of the element 
\be
{\cal P}=
\left(\matrix{0& 0& \ldots & \ldots  & 1\cr
                  1      & 0       & \ldots & \ldots  & 0  \cr
                  0       & 1       &  0  & \ldots  & 0 \cr
                  \ldots      & \ldots     &  \ldots& \ldots  & \ldots\cr
                  0         & 0       & \ldots & 1    & 0 \cr}\right)\0
\ee
of the Weyl group. The boundary condition $\hat X^i(2\pi) = {\cal P} 
\hat X^i(0) {\cal P}^{-1}$ implies that $x^i_k(2\pi) =x^i_{k-1}(0)$, and so
the $x^i_k$ form a unique long string of length $2\pi N$. 

If we believe the limit we have described is the true strong coupling limit,
we can therefore interpret it as the weak coupling limit of (type II) string 
theory, $g_s \sim g^{-1}$,~\cite{motl,DVV}.

\subsection{2D Instantons and Hitchin systems}\label{sec:24}

We look now for classical Euclidean supersymmetric configurations that 
preserve half supersymmetry. To this end we set $\theta=0$ and look 
for solutions of the equations $\delta \theta^\pm=0$, i.e.\ from (\ref{esusy}),
\begin{eqnarray}
&&(\frac{i}{2g^2} F_{w\bar w} +
\frac {1}{2} [X^i,X^j]\tilde\gamma_{ij}) \epsilon^+_c =0,\quad D_w X^i
\gamma_i\epsilon^+_c=0\0 \\
&&(-\frac{i}{2g^2} F_{w\bar w} +\frac {1}{2} [X^i,X^j]\gamma_{ij})\epsilon^-_s=0,
\quad\quad D_{\bar w} X^i \tilde\gamma_i\epsilon^-_s=0\,.\label{2dsusyconf}
\end{eqnarray}
Solutions of these equations that preserve one half supersymmetry are the following
ones. Set $X^i =0$ for all $i$ except two, for definiteness $X^i\neq 0$ for 
$i=1,2$; remark that $\gamma_{12}$ is an antisymmetric $8\times 8$ matrix, and
$\gamma_{12}^2=-1$ and therefore its eigenvalues are $\pm i$ ( moreover
$\tilde \gamma_{12}=\gamma_{12}$). It is easy to show that there exists 
$\epsilon^+$ and $\epsilon^-$, each with four independent components, such that 
\begin{eqnarray}
\gamma_{12} \epsilon^\pm = \pm i\epsilon^\pm, \quad
\gamma_1 \epsilon^+= -i\gamma_2 \epsilon^+,\quad \tilde \gamma_1\epsilon^-=
i\tilde \gamma_2\epsilon^-\,.\0
\end{eqnarray}
Now it is convenient to introduce the complex notation $X=X^1+iX^2$, $~\bar X=
X^1-iX^2= X^\dagger$. Then the conditions to be satisfied in order to preserve
one half supersymmetry are
\begin{eqnarray}
&&F_{w\bar w} + i g^2 [X, \bar X] =0\label{insteq1}\\
&&D_w X=0, \quad\quad D_{\bar w} \bar X=0\,.\label{insteq2}
\end{eqnarray}
It is easy to verify that, if non--trivial solutions to such equations exist,
they satisfy the equations of motion of the action (\ref{eSYM}). The 
action with $\theta=0, X^i=0$ for $i=3,\ldots8$ is
\begin{eqnarray}
S_{inst}=\frac{1}{2\pi} \int d^2w \,\Tr \left(
D_w X D_{\bar w}\bar X +D_w \bar X D_{\bar w} X- \frac{1}{2g^2} F_{w\bar w}^2 +
\frac{g^2}{2}[X,\bar X]^2 \right).\label{instaction}
\end{eqnarray}
It is elementary to prove that $S_{inst}$ vanishes in correspondence to 
solutions of (\ref{insteq1}, \ref{insteq2}) that are single--valued on the 
cylinder.  

From a mathematical point of view, (\ref{insteq1}, \ref{insteq2}) are easily
seen to identify a {\it Hitchin system}~\cite{hitchin} on a sphere with two
punctures (or on an annulus). In such systems, $F$ is the gauge curvature
in reference to a gauge vector bundle $V$, and $\bar X$ is the holomorphic 
section of the bundle $End V\otimes K$, where $K$ is the canonical line
bundle over the base (which is trivial in our case). We would tend to identify 
the moduli space of instanton solutions with the moduli space of the solutions
of the Hitchin systems (which contains the moduli space of Riemann surfaces
with punctures/boundaries). However in the present paper we limit ourselves
to finding explicit sample solutions of 
(\ref{insteq1}, \ref{insteq2}) and show their connection with the affine
Toda field theories.\footnote{Interesting connections between Hitchin
systems and the Toda field theories were previously found in \cite{Aldo}.}
The most general case will be dealt with elsewhere.

But before we end the section let us return, for completeness, to the Minkowski
case. Proceeding as in the Euclidean case, we can find configurations that 
preserve one half supersymmetry. The equations to be satisfied are now
\begin{equation}
F_{+-} + ig^2[X,\bar X] =0, \quad\quad D_- X=0,\quad D_+\bar X =0\,, 
\label{minkinst}
\end{equation}
where we have introduced the light--cone variables 
$\sigma_\pm = \frac{1}{2}(\tau\pm \sigma$),  and $A_\pm = A_0\pm A_1$. 
In the Minkowski case too a configuration satisfying (\ref{minkinst}) 
satisfies the equations of motion of (\ref{mSYM}), and the corresponding value
of the action is zero.

\section{2D instantons and affine Toda systems}\label{sec:3}

Let us consider the BPS equations (\ref{insteq1}, \ref{insteq2})
 and let us define the spectral connection
\be
\A_w=A_w+{\lambda g}\bX,\quad
\quad
\A_{\bar w}=A_{\bar w}-{g\over\lambda}X\,,\label{spectconn}
\ee
where $\lambda$ is a spectral parameter.

We can rewrite the above BPS equations as the zero curvature condition
for such spectral connection
\be
\F_{{\bar w}w}&=&\dbw\A_w-\dw\A_{\bar w}+i\left[\A_{\bar w},\A_w\right]\0\\
&&=\left(F_{{\bar w}w}-{i g^2}\left[X,\bX\right]\right)
+{\lambda g}\left(D_{\bar w}\bX\right)
-{g\over\lambda }\left(D_wX\right)=0\,,\0
\ee
for generic values of the spectral parameter.
The remark that a Hitchin system corresponds to a zero curvature condition
is present in~\cite{hitchin}, although not in the same form as here.

Our purpose now is to find a general ansatz for the solutions of this 
integrable system. In so doing we follow and adapt the
suggestions of~\cite{wynter,GHV}.

First we parametrize the holomorphic component of the connection
as $A_w=-iY^{-1}\dw Y$ and write $X=Y^{-1}MY$, where $Y$ is a generic 
element in the complexified gauge group, i.e.\ $GL(N,C)$. Then the 
equation $D_w X=0$ is equivalent to the equation $\dw M=0$.

Let us define the polynomial
\be
P(\mu)={\rm det}\left(\mu {\bf 1}_N - X\right)
=\mu^N+a_{N-1}\mu^{N-1}+ \ldots +a_0\0
\ee
and notice that $P(M)=0$ by definition. 

To uniquely single out the parametrization, we can choose the matrix $M$ 
to be 
\be
M=\left(\matrix{-a_{N-1}& -a_{N-2}& \ldots & \ldots  & -a_0\cr
		  1      & 0       & \ldots & \ldots  & 0  \cr
		  0       & 1       &  0  & \ldots  & 0 \cr
		  \ldots	   & \ldots     &  \ldots& \ldots  & \ldots \cr		
                  0	    & 0       & \ldots & 1    & 0 \cr
}\right).\label{M}
\ee
Since $\dw M=0$, it follows that $\dw a_i=0$ for each $i=0,\ldots.,N-1$.
Therefore the equation
$P(X)=0$ can be interpreted as the description of an $N$ sheeted 
branched covering of a Riemann surface. The covering map can be obtained by
diagonalizing the $M$ matrix.

With this parametrization, the spectral connection reads
\be
\A_w=-iY^{-1}\dw Y+{\lambda g} Y^\dagger M^\dagger (Y^\dagger)^{-1},
\quad
\A_{\bar w}=i\dbw Y^\dagger (Y^\dagger)^{-1}-
{g\over\lambda} Y^{-1} M Y\,.\label{specc2}
\ee
from which it is easy to extract the zero curvature equation and see that
it is written in terms of $YY^\dagger$ only.

\subsection{The $\Z_N$ coverings}\label{sec:Zn}

 In this subsection we abandon complete generality and
specialize the above structure to the $\Z_N$ covering
$X^N-a=0$, where $\dw a=0$. We will see that the zero curvature system 
reduces in this case to an affine $U(N)$ Toda system.
In this case the $M$ matrix can be rewritten as
\be
M=J^{-1}\left(a^{1/N}{\cal P}\right) J\,,\label{M'}
\ee
where ${\cal P}$ is the permutation matrix of section~\ref{sec:23}, 
and $J={\rm diag}\left(a^{(1-N)/N},a^{(2-N)/N},\ldots,1\right)$.
Let us define the (unitary) matrix of `phases' 
$U=\left(J/{\bar J}\right)^{1/2}$.

A very economical ansatz is obtained by restricting 
the $Y$ field to the complexified Cartan torus. Therefore, taking
\be
Y=e^{ {\beta\over 2} (u\cdot H)} \cdot  \left(J {\bar J}\right)^{-1/2}\,,
\label{Y}
\ee
where $\beta$ is a real coupling and $(u\cdot H)=\sum_i u_i H_i$ is a generic 
field in the Cartan subalgebra of $U(N)$, whose generators $H_i$ are taken to be 
diagonal and hermitean, we get the spectral connection
\be
&&\A_w=U^\dagger\left(-i\dw + \A'_w\right)U,\qquad\qquad
\A_{\bar w}=U^\dagger\left(-i\dbw + \A'_{\bar w}\right)U\0\\
&&\A'_w= -i{\beta\over2}\dw (u\cdot H)
+{\lambda g}~\exp\{{\beta\over 2}(u\cdot H)\}
\,\left({\bar a}^{1/N}{\cal P}^\dagger \right)\exp\{- {\beta\over 2}(u\cdot H)\}
\label{specc1}\\
&&\A'_{\bar w}= i{\beta\over2}\dbw (u\cdot H)
-{g\over\lambda}~\exp\{ -{\beta\over 2} (u\cdot H)\}
\,\left(a^{1/N}{\cal P} \right)\exp\{{\beta\over 2}(u\cdot H)\}\,.\0
\ee
Since the $U$ gauge transformation is singular, we have
\be
\F_{{\bar w}w}=U^\dagger\F'_{{\bar w}w}U-i\left(\dbw\left(U^\dagger\dw U\right)
-\dw\left(U^\dagger \dbw U\right)\right).
\ee
Explicit evaluation gives: $U^\dagger\dw U= \frac{\cal H}{2}\dw 
{\rm ln} \frac{a}{\bar a}$, where
${\cal H}={\rm diag}\left({1-N\over N},{2-N\over N},\ldots ,0\right)$
Therefore, the original zero curvature condition becomes
\be 
\F'_{{\bar w}w}={i\over 2}~\cH~(\dbw\dw -\dw \dbw)\, \ln\left(\frac{a}
{\bar a}\right)= -i\pi \cH~ {\dw a}~ {\dbw \bar a}~
\delta(a)\,.\label{delta}
\ee

To eliminate the $a^{1/N}$ factors, let us transform the coordinates to
$\zeta$ such that ${\partial\zeta\over\partial w}= {\bar a}^{1/N}$.
The spectral connection in the new coordinate system is
\be
&&\A'_\zeta= -i{\beta\over2}\dc (u\cdot H)
+{\lambda g}~\exp\{ {\beta\over 2} (u\cdot H)\}
{\cal P}^\dagger
~\exp\{- {\beta\over 2} (u\cdot H)\} \0\\
&&\A'_{\bar \zeta}= i{\beta\over2}\dbc (u\cdot H)
-{g\over\lambda}~\exp\{ -{\beta\over 2} (u\cdot H)\}
{\cal P}
~\exp\{ {\beta\over 2} (u\cdot H)\}\,, \label{ATFT}
\ee
which we can recognize to be the spectral connection of the affine Toda field 
theory ~\cite{olitu,holli}. From the above discussion it is now clear that 
the resulting equations for the $u\cdot H$ fields are the affine Toda field 
equations with, in addition, a delta--type interaction
at the branching points of the covering, namely
\be
\dc\dbc (u\cdot H)-{ g^2\over \beta}\left[e^{-\beta(u\cdot H)}{\cal P}
e^{\beta(u\cdot H)},{\cal P}^\dagger\right]=
{\pi\over\beta}
{\cal H}\delta(a)(\dbc a)(\dc {\bar a})\,.\label{deltasing}
\ee
The delta--type interaction means a logarithmic boundary condition for a 
combination of the $u_i$'s, say 
\be
u_{\cal H} =  {2\over\beta} \,\ln(|a|)\,.\label{sing}
\ee
That such solutions of the Toda equations exist, has been proven in several 
instances: \cite{cecva,GHV}.

It remains for us to say a few words about the constant $\beta$. It is an 
overall normalization constant for the $u$ fields. By singling it out we can 
connect it to the YM coupling $g$. In the strong coupling limit the BPS 
equations become
\be
F_{w\bar w} =0 ,\quad [X, \bar X] =0,\quad D_wX=0,\quad D_{\bar w}\bar X=0\,,
\ee
i.e.\ the gauge connection is flat and we can diagonalize $X$ and $\bar X$ 
with the same unitary matrix. Comparing with the above we see that 
$\beta \to 0$ as $g\to \infty$.

\section{Interpretation and discussion}\label{sec:4}

Let us summarize what we have done in the previous subsection and 
provide an interpretation of the results.  Our purpose was to parametrize 
solutions of the BPS equations (\ref{insteq1}, \ref{insteq2}) that interpolate 
between different string configurations. We emphasize that the final result, 
$X,\bar X$ and the curvature $F$, must be regular, in particular 
single--valued: what happens is that the 
singularity carried by the Toda fields kills exactly the singularity which 
is responsible for the string interaction, (\ref{sing}). 
In order to see the latter point we have to go to the diagonal picture, 
in which the string interpretation is evident, see section~\ref{sec:23}. 

In what follows we discuss in general the $\Z_N$ covering. A more detailed account can be found 
in the Appendix for the case $N=3$. Let us start by
diagonalizing the matrix $a^{1/N}{\cal P}= \Lambda^{-1} \hat X \Lambda$, where 
$\hat X = a^{1/N}{\rm diag}(1,\ldots,\omega^{N-2},
\omega^{N-1})$ and $\Lambda_{ij}= \omega^{(i-1)(j-1)}$, $\omega$ being 
the primitive $N$--th root of unity,~\cite{wynter}. $\hat X$ is the 
matrix of eigenvalues of $X$, which can now be represented as follows
\be 
X= e^{\frac{\beta}{2} u\cdot H} V^{-1}\hat X V 
e^{-\frac{\beta}{2} u\cdot H}\,, \label{Xnew}
\ee
where $V= \Lambda \sqrt {J/\bar J}$. Now, let us suppose for simplicity that 
$a$ has a simple zero at $z=z_0$ and draw a cut from $z_0$ running in the 
region $|z|>|z_0|$, for simplicity.
Crossing the cut, $V \to {\cal P}^{-1}V$ and $\hat X\to {\cal P}^{-1} \hat X 
{\cal P}$, while all the rest in (\ref{Xnew}) remains unchanged.
Therefore, if we go around the origin of the $z$--plane, $X\to X$ and 
$\hat X\to \hat X$ as long as we do not cross the cut; but, if going 
around the origin we cross the cut, then
$X\to X$, but simultaneously the eigenvalues of $\hat X$ get permuted as
in section~\ref{sec:23}. 

Now notice that, in the strong coupling limit $e^{\frac{\beta}{2} u\cdot H}
\to 1$ and $X \to V^{-1}\hat X V $, and $V$ simultaneously diagonalizes
$X$ and $\bar X$. Therefore $\hat X$ is what remains of $X$ in the strong 
coupling limit of the theory (section~\ref{sec:23}) and can be interpreted in 
terms of string configurations. At this point we can say that
for $|z|<|z_0|$, we have a configurations of $n$ separated 
strings, and for $|z|>|z_0|$ we get a long string obtained by joining the 
previous ones.

Let us see the same problem from the point of view of branched coverings,
and, for definiteness, let us consider covering by means of annuli. 
If the covering were by spheres, we would construct the world--sheet by 
cutting $N$ copies of ${\bf CP_1}$ along lines connecting the zeroes of 
$a$ and joining them in the usual way. Since the covering is by means 
of annuli, the zeros of $a$ which are in the removed disks
are now uneffective up to the fact that some cuts can connect them to
zeros which are in the annulus. This way branch cuts may terminate
on the boundary of each of the annuli and the identification along them
generates the long string configurations. In the case the boundary
is not crossed by cuts, then the related state represents $N$ strings
of lenght one. Each cut internal to the annulus corresponds instead to 
a full joining of the $N$ short strings and then to a total resplitting 
in short ones again. 

In the Appendix we discuss the meaning of other zeroes of $a$ and related 
problems and outline more general coverings.

The conclusion is therefore that the string joining/splitting interactions 
are mediated by 
instantons and geometrically described by suitable (branched coverings of) 
Riemann surfaces, whose genus is determined by the number of distinct zeroes 
of $a$. The branched covering encodes the string world-sheet structure of 
the specific interaction. Each branch represents the world--sheet of a 
string that joins other strings along a cut. The joining is represented by 
the exchange of eigenvalues referred to above,
which in turn corresponds to crossing the branch cuts.

To end this paper let us make a final remark on the possible utilization of
the above results. It was conjectured in~\cite{grome} that the amplitude
for a transition from one string configuration to another be dominated by
the $\Z_N$ coverings we have been considering here. If that is so 
we can therefore try to compute this amplitude by expanding the action around
the appropriate instanton configurations and then summing over them. Since
the instanton action vanishes (this is actually true for any instanton
configuration, see section~\ref{sec:24}), we expect, from the
fluctuations, a result proportional to some power of $g_s$. 
Whether this power corresponds to the one required by 
perturbative string field theory, will be a crucial test for the compatibility 
of the latter with MST.

\acknowledgments
We would like to acknowledge valuable discussions we had 
with E.\ Aldrovandi, D.\ Amati, M.\ Bertolini, B.\ Dubrovin, R.\
Iengo, T.\ Grava and J.B.\ Zuber. We would
like to thank in particular C.S.\ Chu for his collaboration in the
early stage of this work. This research was partially supported by EC
TMR Programme, grant FMRX-CT96-0012, and by the Italian MURST for the
program ``Fisica Teorica delle Interazioni Fondamentali".

\appendix

\section{$\Z_3$ covering}

We present here as explicitly as possible some solutions relative to the 
$N=3$ case. For the $\Z_3$ covering, these are described by the following  
spectral equation: 
\be X^3 = a(\bar z) \ee
where $X$ is the complex $3\times 3$ matrix, and $a$ is antiholomorphic. 
The matrix $M$ which solves this equation has the following 
form:\footnote{It is shown in diagonal form using ($\o=\e^{2\pi i/3}$):
		\be \cP=\bma0&0&1\\1&0&0\\0&1&0\ema = \Lm^{-1} \bma1\\&\o\\&&\o^2\ema\Lm\qquad
		\Lm={1\over\sqrt{3}}\bma1&1&1\\1&\o&\o^2\\1&\o^2&\o\ema\qquad
		J=a^\cH=\bma \!a^{-2/3}\\&\!\!a^{-1/3}\\&&\!\!1\ema\,.
		\ee}
\be M &=&  \bma0&0&a\\1&0&0\\0&1&0\ema = a^{1/3}J^{-1}\cP J
  =a^{-\cH}\Lm^{-1}\left[a^{1/3}\bma1\\&\o\\&&\o^2\ema\right]\Lm a^\cH\,,
\ee
This shows, with the parameterization of section~\ref{sec:Zn}, the
full string matrix $X$ in its diagonal and gauge parts: 
\be X &=& \left({a\over \bar a }\right)^{-\frac\cH2} 
	\e^{-\beta u\cdot H}\Lm^{-1}
	\left[a^{1/3}\bma1\\&\o\\&&\o^2\ema\right]\Lm
	\left({a\over \bar a }\right)^\frac\cH2 \e^{\beta u\cdot H}=\\
&=& \e^{-(\frac12\ln{a\over\bar a}\cH+\beta u\cdot H)}\Lm^{-1}
    \left[a^{1/3}\bma1\\&\o\\&&\o^2\ema\right]
    \Lm\e^{-(\frac12\ln{a\over\bar a}\cH+\beta u\cdot H)}\,.
\ee
Notice that $\frac12\ln(a/\bar a)$ is purely imaginary being the
phase of $a$, while the functions $u$ are real. The two pieces, then,
represent the unitary and the hermitean parts of the complexified gauge
transformation which diagonalizes $X$.

To show that $X$ is indeed single-valued, it is sufficient to pass the
$\Lm$ factor through the Cartan generators $H'=\Lm H\Lm^{-1}$, 
$\cH'=\Lm\cH\Lm^{-1}$ and notice that $\exp(-2i\pi\cH')=\cP$, the
permutation matrix. Then
\be X = \Lm^{-1}\e^{-\beta u\cdot H'}\,
	\cP^{{1\over4\pi i}\ln{a\over\bar a}}\,
	\left[a^{1/3}\bma1\\&\o\\&&\o^2\ema\right]\,
	\cP^{-{1\over4\pi i}\ln{a\over\bar a}}\,
	\e^{-\beta u\cdot H'}\Lm\,.
\ee
Around points where $a$ has simple zeroes, and its phase passes from $0$
to $2\pi$, the exponent of $\cP$ passes from $0$ to $1$, thus giving the
permutation matrix. This latter rotates the eigenvalues, compensating 
the $\o$ factor which appears from the $a^{1/3}$. 
This mechanism works also for generic zeroes or poles of $a$.

\bigskip
As an example, using the coordinate on the annulus  $z=\e^w$, let's
take $\bar a=z-z_0=|a|\e^{i\phi}$. This function describes the joining of
three short strings in one long string of length 3 (it gives in fact a single
branch of order three on the triple covering of the cylinder).
The interaction takes place at $z_0$. We can explicitly write the solution:
\be 
X = \Lm^{-1}\e^{-\beta u\cdot H'} 
	\cP^{-{\phi\over2\pi}}
	\left[\sqrt[3]{\bar z-\bar z_0}\bma1\\&\o\\&&\o^2\ema\right]
	\cP^{\phi\over2\pi}
	\e^{-\beta u\cdot H'}\Lm\,.
\ee
On the complex $z$ plane the eigenvalues of $X$ have a branch cut of order 3,
originating at $z_0$ and extending to infinity, i.e.\ to the end of the
cylinder. At the same time their monodromy is determined by the winding
around $z_0$ of any path centered at the origin (round trips around
the cylinder). It is easy to see that the monodromy changes from the
identity to $\cP$, the permutation matrix.

\EPSFIGURE{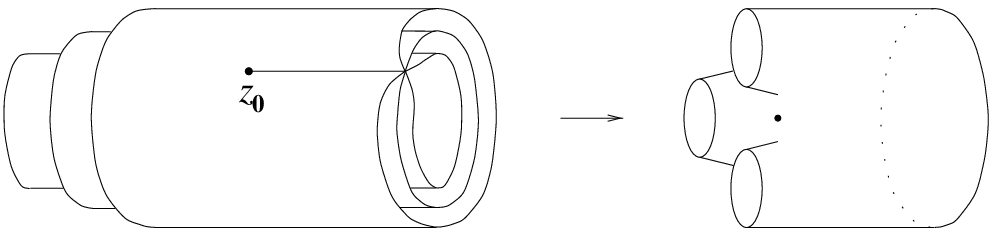}{Joining of three short strings together}

The functions $u_i$ have to be determined from the Toda/self-duality
equation~(\ref{deltasing}), which in the $z$ coordinates reads: 
\be \d_z\d_{\bar z} (u\cdot H)
-{g^2\over \beta}{|z-z_0|^{2/3}\over|z|^2}
\left[e^{-\beta(u\cdot H)}\cP e^{\beta(u\cdot H)},\cP^\dagger\right]=
{\pi\over\beta}\cH\delta^{(2)}(z-z_0)\,.
\ee

This equation requires for some field $u_\cH$ related to the direction
$\cH$ in Cartan space, a particular boundary condition at $z=z_0$,
namely: 
\be u_\cH \simeq \frac2\beta \ln(|z-z_0|)\,.\ee

\bigskip
Other choices of the function $a$, with more than one zero, can be
constructed to give subsequent interactions, possibly increasing 
the genus of the world--sheet. For example, let us consider the case $N=2$ 
with $\bar a$ having two zeroes inside the annulus. When
the branch cuts are chosen to exit from different boundaries, the
covering describes a 2-string to 2-string process of genus 1. 
On the other hand the case in which the two cuts exit from the same
boundary, or the case in which there is a single branch cut joining the zeroes, 
corresponds to a genus 0 scattering of two short strings.

It is clear that in order to describe more general string 
configurations one should consider more general branched covering
structures. For example, if an integral factorization $N=LM$ exists,
then the covering structure $\left(X^M-a\right)^L+b=0$ can describe
splitting and joining of $L$ strings of length $M$ in $N$ lenght one 
strings and in one lenght $N$ string.

\end{document}